\documentclass[pra,twocolumn,showpacs,preprintnumbers]{revtex4}
\usepackage{graphicx}
\usepackage{amsfonts}
\usepackage{amssymb}
\usepackage{amsmath}
\makeindex
\newcommand{\be}{\begin{equation}}
\newcommand{\ee}{\end{equation}}
\newcommand{\bea}{\begin{eqnarray}}
\newcommand{\eea}{\end{eqnarray}}
\newcommand{\Eq}[1]{Eq.\,(\ref{#1})}
\newcommand{\Fig}[1]{Fig.\,\ref{#1}}
\newcommand{\Sec}[1]{Sec.\,\ref{#1}}
\newcommand{\Onlinecite}[1]{Ref.\,\onlinecite{#1}} 

\newcommand{\GFk}{\hat{\mathbf{G}}_k}
\newcommand{\GFkm}{\hat{\mathbf{G}}_{k_\mu}}
\newcommand{\GFkOD}{G_k}
\newcommand{\GFkfs}{\hat{\mathbf{G}}^{fs}_k}

\newcommand{\br}{\mathbf{r}}
\newcommand{\bk}{\mathbf{k}}

\newcommand{\bE}{\mathbf{E}}

\newcommand{\En}{\mathbf{E}_n}
\newcommand{\Hn}{\mathbf{H}_n}
\newcommand{\EnOD}{E_n}
\newcommand{\An}{\mathbf{A}_n}
\newcommand{\AnOD}{A_n}

\newcommand{\brd}{\mathbf{\dot{r}}}

\newcommand{\heps}{\hat{\boldsymbol{\varepsilon}}}

\newcommand{\hepsOD}{\varepsilon}
\newcommand{\hsigma}{\hat{\boldsymbol{\sigma}}}

\begin{document}
\title{Resonant-state-expansion Born approximation with a correct eigen-mode normalisation}
\author{ M.\,B. Doost}
\affiliation{Independent Researcher, United Kingdom}
\begin{abstract}
The Born approximation (Born 1926 Z.Phys.38.802) is a fundamental result in physics, it allows the calculation of weak scattering via the Fourier transform of the scattering potential. As was done by previous authors 
(Ge {\it et al} 2014 New J. Phys. 16 113048) the Born approximation is extended by including in the formula the resonant-states (RSs) of the scatterer. However in this study unlike previous studies the included eigen-modes 
are correctly normalised with dramatic positive consequences for the accuracy of the method. The normalisation of the RSs used in the previous RSE Born approximation or resonant-state-expansion Born 
approximation made in Ge {\it et al} (2014 New J. Phys. 16 113048) has been shown to be numerically unstable in Muljarov {\it et al} (2014 arXiv:1409.6877) and by analytics here. The RSs of the system can be calculated using 
my recently discovered RSE perturbation theory for dispersive electrodynamic scatterers (Muljarov {\it et al} 2010 Europhys. Lett. 92 50010; Doost {\it et al} 2012 Phys, Rev. A 89; Doost {\it et al} 2014 Phys. Rev. A 90 013834) and 
normalised correctly to appear in the spectral Green's functions and hence the RSE Born approximation via the flux-volume normalisation which I recently rigorously derived in Armitage {\it et al} (2014 Phys. Rev. A 89), 
Doost {\it et al} (2014 Phys. Rev. A 90 013834)(2016 Phys. Rev. A 93 023835). In the case of effectively one-dimensional systems I find an RSE Born approximation alternative to the scattering matrix method.
\end{abstract}
\date{\today}
\maketitle
\section{Introduction}

Fundamental to scattering theory, the Born approximation consists of taking the incident field in place of the total field as the driving field at each point inside the scattering potential, 
it was first discovered by Max Born and presented in  Ref.~\cite{Born26}. The Born approximation gave an expression for the differential scattering cross section in terms of the Fourier transform 
of the scattering potential. The Born approximation is only valid for weak scatterers as we will see in the numerical demonstrations. 

In this paper I provide an extension to the Born approximation which allows an arbitrary number of resonant states (RSs) to be taken into account. I have named this extension to the Born approximation 
the Resonant-state-expansion correction to the Born approximation or the RSE Born approximation. An almost identical approach is already available in the literature \cite{Ge14} however its derivation differed
by including an unstable normalisation formula for the RS eigen-modes of the system which was then subsequently used to expand Born's approximation incorrectly. The normalisation derived in 
Ref.\cite{LeungPRA94} and used in the previous RSE Born approximation made in Ref.\cite{Ge14} has been shown to be numerically unstable in Ref.\cite{MuljarovARX14} and shown to be unstable using analytics in Appendix C. Furthermore 
the numerical study made in Ref.\cite{Ge14} only included a single RS in the expansion of the Born approximation, most likely to avoid divergence caused by their incorrect normalisation of the RSs.

Recently there has been developed~\cite{MuljarovEPL10, NMuljarovEPL101, NMuljarovEPL102} a rigorous perturbation theory called resonant-state expansion 
(RSE) which was then applied to one-dimensional (1D), 2D and 3D 
systems~\cite{DoostPRA12,DoostPRA13,ArmitagePRA14,DoostPRA14,DoostARX15A,DoostARX15B} which only calculates the modes and 
makes no use of them. The RSE accurately and efficiently calculates RSs of an arbitrary system in 
terms of an expansion of RSs of a simpler, 
unperturbed one. RSs are normalised correctly to appear in spectral Green's functions (GFs) via the flux volume 
normalisation\cite{DoostPRA14} and hence the RSE Born approximation. In the limit where 
an infinite number of these resonances are included in the RSE Born approximation we will observe convergence of 
the method towards the exact solutions. That the RSE can reproduce both the correctly normalised
RS fields as well as frequencies was demonstrated in Ref.\cite{DoostPRA12} with the convergence and extrapolation 
algorithm which I contributed to that paper.

Interestingly the resonant-state-expansion is a near identical translation to Electrodynamics 
of a much earlier theory from Quantum Mechanics by More, Gerjuoy, Bang, Gareev, Gizzatkulov 
and Goncharov Ref.\cite{NMuljarovEPL101, NMuljarovEPL102}. The only difference between the 
two approaches is the choice of RS normalisation method.
I am able
to show in this manuscript that the general normalisation of resonant-states which I derived in Ref.\cite{DoostPRA14} is the most numerically 
stable available normalisation method. The general normalisation which I derived in Ref.\cite{DoostPRA14} is based on a prototype normalisation which appeared in 
Ref.\cite{MuljarovEPL10}.

The concept of RSs was first conceived and used by Gamow in 1928 in order to describe mathematically the process of radioactive decay, specifically the escape from the nuclear potential of an alpha-particle 
by tunnelling. Mathematically this corresponded to solving Schr\"odinger's equation for outgoing boundary conditions (BCs). These states have complex frequency $\omega$ with negative imaginary part meaning 
their time dependence $\exp(-i\omega t)$ decays exponentially, thus giving an explanation for the exponential decay law of nuclear physics. The consequence of this exponential decay with time is that the 
further from the decaying system at a given instant of time the greater the wave amplitude. An intuitive way of understanding this divergence of wave amplitude with distance is to notice that waves that are 
further away have left the system at an earlier time when less of the particle probability density had leaked out. There already exists numerical techniques for finding eigenmodes such as finite element 
method (FEM) and finite difference in time domain (FDTD) method to calculate resonances in open cavities. However determining the effect of perturbations which break the symmetry presents a significant 
challenge as these popular computational techniques need large computational resources to model high quality modes. Also these methods generate spurious solutions which would damage the accuracy of the 
RSE Born approximation if included in the basis.

The paper is organized as follows, \Sec{Sec:RSE1} outlines the derivation of the RSE Born approximation, \Sec{Sec:Alternative normalisations} discusses normalisations of RSs by other authors, \Sec{sec:Unperturbed} 
outlines the application of the RSE Born approximation to planar slabs, 
\Sec{sec:application3D} gives the numerical validation of the new method along with a comparison of the alternative RSE approaches.  
\section{Derivation of the RSE Born approximation}
\label{Sec:RSE1}

I will in the following section re-derive the method for calculating the full GF of an open electrodynamic system in the same way as Ref.\cite{Ge14} however unlike previous authors I 
use the numerically stable normalisation of
RSs which I derived in Ref.\cite{DoostPRA14,DoostARX15B}. These methods are required to calculate transmission and scattering cross-section from the dispersive RSE perturbation theory with mathematical rigour and 
accuracy.

For an electrodynamic system with local frequency dependent dielectric permittivity tensor $\heps_k(\br$) and permeability $\mu = 1$, where $\br$ is the three-dimensional spatial position, 
Maxwell's wave equation for the electric $\mathbf{E(\br)}$-field with a current source $\mathbf{J(\br)}$ oscillating at frequency $k$, which can be real or complex, is 
 \be
- \nabla\times\nabla\times\mathbf{E(\br)}+k^2\heps_k(\br)\mathbf{E(\br)}=ik\dfrac{4\pi}{c}\mathbf{J(\br)}\,.
 \label{EJequ0}
\ee
The time-dependent part of the field is given by $\exp(-i\omega t)$ with the 
complex eigen-frequency $\omega=c k$, where $c$ is the speed of light in vacuum.

The Green's function (GF) of an open electromagnetic system is a tensor $\GFk$ which satisfies Maxwell's wave equation \Eq{EJequ0} with a delta function source term,
\be
- \nabla\times\nabla\times \GFk(\br,\br')+k^2\heps_k(\br)\GFk(\br,\br')=\hat{\mathbf{1}}\delta(\br-\br')\,,
 \label{GFequ00}
\ee
where $\hat{\mathbf{1}}$ is the unit tensor. Physically, the GF describes the response of the system to a point current with frequency $\omega$, i.e. an oscillating dipole.

The importance of $\GFk$ comes from the fact we can see from Eqs.\,(\ref{GFequ00}) that Eqs.\,(\ref{EJequ0}) can be solved for $\mathbf{E(\br)}$ by convolution of $\GFk$ with the current source $\mathbf{J(\br')}$,
\be 
\mathbf{E(\br)}=\int\GFk(\br,\br')ik\dfrac{4\pi}{c}\mathbf{J(\br')} d\br'\,.\label{GEA1equ}
\ee
Inside the system we can use the RSE to calculate the GF. In Appendix A I derive for dispersive systems (for which I have recently developed a dispersive RSE perturbation theory \cite{DoostARX15A,DoostARX15B}) 
a convenient form of the spectral GF, valid inside the scatterer only,
\be \GFk(\br,\br')=\sum_n \frac{\En(\br)\otimes\En(\br')}{2k(k-k_n)}\,.
\label{ML4} \ee
The $\En$ are RSs of the open optical system and are defined as the eigen-solutions of Maxwell's wave equation,
\be
\label{me3D}
\nabla\times\nabla\times\En(\br)=k_n^2\heps_k(\br)\En(\br)\,,
\ee
satisfying the {\it outgoing wave} BCs. I have also taken the resonator to be embedded in free space ($\heps=1$) without loss of generality. Here, $k_n$ is the wave-vector eigen-value of the RS numbered by the index 
$n$, and $\En(\br)$ is its electric field eigen-function. The RSs which are 
solutions of \Eq{me3D} which are either stationary or decaying in time.
Modes appearing in \Eq{ML4} are normalized \cite{DoostPRA14,DoostARX15B} according to the flux-volume normalisation
\bea
\label{normaliz}
\delta_{0,k_n}+1&=&\int_V\En(\br)\cdot\dfrac{\partial(k^2\heps_k(\br))}{\partial(k^2)}\bigg|_{k=k_n}\En(\br)d{\bf r}\\
&&+\lim_{k\rightarrow k_n}\oint_{S_V} \dfrac{\En\cdot\nabla\bE-\bE\cdot\nabla\En}{k^2-k^2_n} d{\bf S}\,,\nonumber
\eea
where the first integral is taken over an arbitrary simply connected volume $V$ enclosing the inhomogeneity of the system and the center of 
the spherical coordinates used, and 
the second integral is taken over its surface $S_V$. This normalization is required \cite{DoostPRA14} for the 
validity of the spectral representation \Eq{ML4}. Numerically \Eq{normaliz} has been validated by its use in RSE perturbation theory 
~\cite{DoostPRA12,DoostPRA13,ArmitagePRA14,DoostPRA14,DoostARX15A,DoostARX15B}. A discussion of the dispersive RSE for nano-particles is 
given in Appendix B.

The generalisation of my \Eq{normaliz} to $k_n=0$ modes is attributable solely to E. A. Muljarov in Ref.\cite{DoostPRA14} 
(I derived the normalisation proof for Ref.\cite{DoostPRA14} without $k=0$ modes), 
however that part of the proof of 
the normalisation can only be further 
generalised to dispersive systems using the spectral GF \Eq{ML4} derived in Appendix A. 
The required derivation is identical except that it makes use of the rigorously derived spectral GF \Eq{ML4}
instead of the identical GF derived in a less mathematically rigorous way for non-dispersive systems. 
This last step is vital for the accuracy of the method. Further I note that just as I explained in Ref.\cite{DoostARX15B}
$\heps_k$ must be a symmetric matrix or a scalar in order to calculate the dispersion factor as shown. 

Various schemes exist to evaluate the surface integral limit in \Eq{normaliz} such as analytic methods in 
Ref.\cite{MuljarovEPL10,DoostPRA14,MuljarovARX14} or numerically extending the surface into a non-reflecting, absorbing, perfectly matched layer where it vanishes.  

The derivation of the RSE Born approximation by Ge {\it et al}.~\cite{Ge14} has been made using the 
normalization introduced by Leung {\it et al.}~\cite{LeungPRA94} the limit of infinite volume ${V}$ is taken:
\be
1=\lim_{{V}\to\infty} \int_{{V}} \dfrac{\partial k^2\heps_k(\br)}{\partial(k^2)}\bigg|_{k=k_n} \En^2 (\br) d{\bf r} + \frac{i}{2k_n} \oint_{S_{V}} \En^2 (\br) dS\,,
\label{norm2}
\ee
It was numerically found \cite{KristensenOL12} that the surface term was leading to a stable value of the integral for the relatively small volumes available in 2D finite difference in time domain (FDTD) 
calculations. However, it was discovered at the time that this was not the case for low-Q modes. It was wrongly shown by Muljarov {\it et al} \cite{MuljarovARX14} 
that \Eq{norm2} is actually diverging in the limit ${V}\to\infty$, and therefore the expansion of the Born approximation in \cite{Ge14} and the normalization \Eq{norm2} are incorrect. In Appendix C
I provided a mathematically rigorous disproof of some of E. A. Muljarov's points and make some correct points about the unsuitability of \Eq{norm2} for the RSE perturbation theory myself. Hence although being a 
cornerstone of the scattering theory of open systems the correct expansion of the Born approximation in terms of RSs to the exact solution was not previously available.

Analogously to Ref.\cite{Ge14} the derivation of the RSE Born approximation of Ge {\it et al}.~\cite{Ge14} is made but in this case using my correct normalisation formula for modes.

That the $\En(\br)$ and $k_n$ can be calculated accurately by the RSE perturbation theory and normalised correctly by \Eq{normaliz} makes possible the RSE Born approximation.

The free space GF $\GFkfs$ is now introduced
\be
- \nabla\times\nabla\times \GFkfs(\br,\br')+k^2\GFkfs(\br,\br')=\hat{\mathbf{1}}\delta(\br-\br')\,,
 \label{GFequ0fs}
\ee
which has the solution,
\be
\GFkfs(\br,\br')=-\dfrac{e^{ik|\br-\br'|}}{4\pi|\br-\br'|}\hat{\mathbf{1}}\,,
 \label{GFequ0fsSol}
\ee

The systems associated with $\GFk$ and $\GFkfs$ are related by the Dyson Equations perturbing back and forth with $\Delta\heps_k(\br)=\heps_k(\br)-\hat{\mathbf{1}}$ \cite{Ge14},
\bea \GFk(\br,\br'')&=&\GFkfs(\br,\br'') \label{Dyson1}\\
&&-k^2\int \GFkfs(\br,\br''')\Delta\heps_k(\br''')\GFk(\br''',\br'') d\br'''\,, \nonumber \eea
\bea \GFk(\br''',\br'')&=&\GFkfs(\br''',\br'') \label{Dyson2}\\
&&-k^2\int \GFk(\br''',\br')\Delta\heps_k(\br')\GFkfs(\br',\br'') d\br'\,, \nonumber \eea
Combining \Eq{Dyson1} and \Eq{Dyson2} it is obtained as in Ref.\cite{Ge14}
\begin{multline}\label{eq:Mark0}
\GFk(\br,\br'')=\GFkfs(\br,\br'')
\\
-k^2\int \GFkfs(\br,\br')\Delta\heps_k(\br')\GFkfs(\br',\br'') d\br'
\\
+k^4\int\int \GFkfs(\br,\br')\Delta\heps_k(\br')\GFk(\br',\br''')
\\
\times\Delta\heps_k(\br''')\GFkfs(\br''',\br'')d\br'''d\br'\,.
\end{multline}
In order to improve the numerical performance further I make a final few steps as in the original Born approximation \cite{Born26}, I define unit vector $\hat{\br}$ such 
that $\br=r\hat{\br}$ and $\bk_s=k\hat{\br}$. Then for $r>>r'$, 
\be
k|\br-\br'|= kr\pm\bk_s\cdot\br'+{\bf O}\left(\dfrac{1}{r}\right)>0
\label{ffa}
\ee
Therefore substituting \Eq{ML4} and \Eq{GFequ0fsSol} in to \Eq{eq:Mark0} and using \Eq{ffa} because both $\br,\br''$ are far from the scatterer we arrive at the RSE Born approximation
\begin{multline}\label{eq:Mark1}
\GFk(\br,\br'')=-\dfrac{e^{ik|\br-\br''|}}{4\pi|\br-\br''|}\hat{\mathbf{1}}
\\
-k^2\dfrac{e^{ik(r+r'')}}{16\pi^2 rr''}\int e^{i(\bk_s-\bk_s'')\cdot\br'}\Delta\heps_k(\br')d\br'
\\
+k^3\dfrac{e^{ik(r+r'')}}{16\pi^2rr''}\sum_n \frac{\An({\bk_s})\otimes\An(-\bk_s'')}{2(k-k_n)}\,.
\end{multline}
or using \Eq{TUES_NIGHT1} instead
\begin{multline}\label{eq:Mark1}
\GFk(\br,\br'')=-\dfrac{e^{ik|\br-\br''|}}{4\pi|\br-\br''|}\hat{\mathbf{1}}
\\
-k^2\dfrac{e^{ik(r+r'')}}{16\pi^2 rr''}\int e^{i(\bk_s-\bk_s'')\cdot\br'}\Delta\heps_k(\br')d\br'
\\
+k^4\dfrac{e^{ik(r+r'')}}{16\pi^2rr''}\sum_n \frac{\An({\bk_s})\otimes\An(-\bk_s'')}{2k_n(k-k_n)}\,.
\end{multline}
The vector $\An$ is defined as a Fourier transform of the RSs,
\be
\An(\bk_s)=\int e^{i\bk_s\cdot\br'}\Delta\heps_k(\br')\En(\br')d\br'\,.
 \label{Ani}
\ee 
I note that the fast Fourier transform method is available. Furthermore I note that for the inverse scattering
problem at resonance the inverse Fourier transformation is also available. 
The first two terms in \Eq{eq:Mark1} correspond to the standard Born approximation, the final summation term corresponds to the 
RSE correction to the Born approximation.

A simple corollary of this theory is as follows, we can see from the arguments just stated that from Eq.(10) if $\br''$ is inside the resonator and $r>>r''$ then
\begin{multline}\label{eq:Mark3b}
\GFk(\br,\br'')=-\dfrac{e^{ik|\br-\br''|}}{4\pi|\br-\br''|}\hat{\mathbf{1}}
\\
+k\dfrac{e^{ikr}}{4\pi r}\sum_n \frac{\An(\bk_s)\otimes\En(\br'')}{2(k-k_n)}\,,
\end{multline}
or using \Eq{TUES_NIGHT1} instead
\begin{multline}\label{eq:Mark3}
\GFk(\br,\br'')=-\dfrac{e^{ik|\br-\br''|}}{4\pi|\br-\br''|}\hat{\mathbf{1}}
\\
+k^2\dfrac{e^{ikr}}{4\pi r}\sum_n \frac{\An(\bk_s)\otimes\En({\br}'')}{2k_n(k-k_n)}\,,
\end{multline}
similarly from Eq.(11) if $\br$ is inside the resonator and $r''>>r$ then
\begin{multline}\label{eq:Mark2}
\GFk(\br,\br'')=-\dfrac{e^{ik|\br-\br''|}}{4\pi|\br-\br''|}\hat{\mathbf{1}}
\\
+k\dfrac{e^{ikr''}}{4\pi r''}\sum_n \frac{\En({\br})\otimes\An(-\bk_s'')}{2(k-k_n)}\,,
\end{multline}
or using \Eq{TUES_NIGHT1} instead
\begin{multline}\label{eq:Mark2}
\GFk(\br,\br'')=-\dfrac{e^{ik|\br-\br''|}}{4\pi|\br-\br''|}\hat{\mathbf{1}}
\\
+k^2\dfrac{e^{ikr''}}{4\pi r''}\sum_n \frac{\En({\br})\otimes\An(-\bk_s'')}{2k_n(k-k_n)}\,,
\end{multline}
other permutations are possible.

\section{Other normalisations}
\label{Sec:Alternative normalisations}
\subsection{Normalisation by Sauvan and co-workers}

The rigorously derived normalisation of Sauvan and co-workers that they gave in Ref.\cite{Sauvan} as 
\bea
\label{normalizCV}
2&=&\int_V\En(\br)\cdot\dfrac{\partial(k^2\heps_k(\br))}{\partial(k^2)}\bigg|_{k=k_n}\En(\br)d{\bf r}\\
&&-\int_V\Hn(\br)\cdot\dfrac{\partial(k^2\heps_k(\br))}{\partial(k^2)}\bigg|_{k=k_n}\Hn(\br)d{\bf r}\,,\nonumber
\eea
requires that the integral $V$ be continued into a perfectly matched layer where it is attenuated to zero, 
thus eliminating the need for surface terms in the normalisation. As such it is most suitable
for FEM and FDTD calculations. Further I note that just as I
explained in Ref.[10] $\heps_k(\br)$ must be a symmetric matrix
or a scalar in order to calculate the dispersion factor as
shown. 

I generate this normalisation by combining the RSE normalisation for $\Hn$ and $\En$, 
I note that this alternative approach to deriving Sauvan's normalisation was discussed with 
E. A. Muljarov at some point but without discussing $k_n=0$ modes.
I also note that $k_n=0$ modes have only ${\bf H}$ field or {\bf E} field component by Maxwell's 
equations because they are curl free modes. $k_n=0$ are by definition modes which satisfy the condition of being curl free. 
These two points explain why the addition of $k_n=0$ modes to Sauvan and co-worker's normalisation takes the form it does. Actually the RSE 
normalisation for $\Hn$ modes was first shown to me in an email attachment, by E. A. Muljarov several years ago but without any 
derivation and without $k_n=0$ modes or differential dispersion factor included.

The rigorous derivation of the relationship between normalised $\mathbf{E}$-field and normalised $\mathbf{H}$-field can be found 
in appendix D of my PhD thesis \cite{Thesis}.

Clearly as the perturbation is increased there is a critical perturbation strength at which the RSE becomes less 
efficient than  FDTD and FEM and beyond this point one should use the RSE Born approximation with
the normalisation of Ref.\cite{Sauvan} and FDTD or FEM.

\subsection{Radiation mode normalisation}

I have recently written a paper on the RSE Born approximation for waveguides with dispersion \cite{DoostARX15B}. 
I found that such modes for cylindrical/effectively-2D waveguides can be normalised by reducing Maxwell's 
equation to effectively 2D and replacing the operation $\nabla\times\nabla\times$ with a suitable linear operator $L$ 
invariant along the length of the waveguide. A similar approach is found in Ref.\cite{Vial} and further
comparison of the two methods is required.

\section{Application to planar systems}
\label{sec:Unperturbed}

In this section we discuss the application of the RSE Born approximation to exactly solvable 1D scattering problems in electrodynamics. 
This is in order to prove the converges of the new method to 
the exact solutions available for 1D problems in \Sec{sec:application3D}. The dielectric profile is described by a scalar frequency 
independent dielectric profile, i.e. 
$\heps_{k}(z)=\hat{\mathbf{1}}\varepsilon(z)$, $\Delta\heps_{k}(z)=\hat{\mathbf{1}}\Delta\varepsilon(z)$. As unperturbed system we 
use a homogeneous planar slab of half width $a$, so that
\be
\varepsilon(z)=\left\{
\begin{array}{lc}  \epsilon_s  & \mbox{for}\ \  |z|<a\,,\\
1 & \mbox{elsewhere}\,.\end{array}\right.
 \label{unpslab}
\ee

\subsection{Wave equation and normalisation formula in 1D}

In this sub-section I consider how Maxwell's wave equation transforms to 1D. I also consider how the normalisation formula transforms to 1D.

Maxwell's wave equations for a planar dielectric structure with permeability $\mu=1$ surrounded by
vacuum is reduced for 1D to the following equation:
\begin{equation}
\partial^{2}_{z}{\mathbb E}_{n}(z,t)=\varepsilon(z){\partial^{2}_{t}{\mathbb E}_{n}(z,t)}\,,
\label{MXW}
\end{equation}
We take the transverse eigen-modes with index $n$ to have
zero in-plane wave number. The eigen-modes can be factorised as
\begin{equation}
{\mathbb E}_{n}(z,t)=\EnOD(z)\exp(-ick_{n}t)\hat{\bf y}
\end{equation}
with time independent part satisfying the wave equation:
\begin{equation}
\Bigl\{\partial^{2}_{z}+\varepsilon(z)k_{n}^2\Bigr\}\EnOD(z)=0\,,
\label{pslab7}
\end{equation}
The electric field and its first derivative are continuous everywhere. Eigenmodes of Maxwell's wave equation for open systems
have outgoing boundary conditions.

In 1D non-dispersive systems the RSs $E_{n}(z)$ with frequency $k_n$
are orthogonal and normalized correctly in 1D according to \cite{MuljarovEPL10}
\begin{eqnarray}
&&\int_{-a}^{a} \varepsilon(z){E}_{n} (z){E}_{m} (z)\,dz \nonumber\\
&&- \frac{{E}_{n} (-a){E}_{m} (-a) +{E}_{n} (a){E}_{m} (a)}{i({k}_{n}+{k}_{m})}=\delta_{nm}\,,
\label{nint}
\end{eqnarray}
where $z=\pm a$ are the positions of the boundaries of the unperturbed system.

\subsection{Resonant states of the unperturbed slab}

In this sub-section I give the RSs used to calculate the RSE Born approximation in \Sec{sec:application3D}.

Solving the wave equation Eq.\,(\ref{pslab7}) for
dielectric constant $\varepsilon(z)$ given by Eq.\,(\ref{unpslab}), the electric field of RS $n$,
normalized according to Eq.\,(\ref{nint}), takes the form \cite{MuljarovEPL10}
\begin{equation}
E_n(z)=\left\{
\begin{array}{lll}
(-1)^nA_ne^{-ik_nz}\,, & & z<-a\,,\\
B_n[e^{i\sqrt{\epsilon_s}k_nz}+(-1)^ne^{-i\sqrt{\epsilon_s}k_nz}]\,, &  &|z|\leq a\,,\\
A_ne^{ik_nz}\,, && z>a\,,
\end{array} \right.
\label{basise1}
\end{equation}
where
\begin{equation}
A_n=\frac{e^{-ik_na}}{\sqrt{a(\epsilon_s-1)}}\,,\qquad B_n=\frac{(-i)^n}{2\sqrt{a\epsilon_s}}\,,
\label{basise2}
\end{equation}
with
\begin{equation}
k_n=\frac{1}{2a\sqrt{\epsilon_s}}(\pi n-i\ln\gamma),\qquad n=0,\,\pm1,\,\pm2,\,...\,, \label{basise3}
\end{equation} and
\begin{equation}
\gamma=\frac{\sqrt{\epsilon_s}+1}{\sqrt{\epsilon_s}-1}\,, \label{basise4}
\end{equation}
the imaginary part of the wave vectors $k_n$ are all the same.

\subsection{The form of the RSE Born approximation in the one dimensional case}

It is demonstrated in this section that the 1D RSE Born approximation can be used in conjuncture with the RSE perturbation theory (to generate the normalised eigen-modes of planar systems with arbitrary dielectric profile and dispersion) 
\cite{MuljarovEPL10,DoostPRA12, ArmitagePRA14, DoostARX15B} to offer a possible alternative to the scattering matrix method of Ref.\cite{CoIng}. The same method for planar waveguides can be developed in an analogous way except
the eigen-modes should be calculated as in Ref.\cite{ArmitagePRA14, DoostARX15B}.

In 1D the GF $\GFkOD(z,z'')$ is the solution of the equation
\begin{equation}
\Bigl\{\partial^{2}_{z}+\varepsilon_k(z)k^2\Bigr\}\GFkOD(z,z'')=\delta(z-z'')\,,
\label{pslab}
\end{equation}
which from \Eq{ML4} we can see is given by
\be
\GFkOD(z,z'')=\sum_n\dfrac{\EnOD(z)\EnOD(z'')}{2k(k-k_n)}\,.\label{ML4OD}
\ee
The free space GF is a solution of 
\begin{equation}
\Bigl\{\partial^{2}_{z}+k^2\Bigr\}\GFkOD(z,z'')=\delta(z-z'')\,,
\label{pslab}
\end{equation}
and is given by 
\be
\GFkOD(z,z'')=-\dfrac{e^{ik|z-z''|}}{2ik}\,.
\ee

Hence in 1D the RSE Born approximation is greatly simplified to
\begin{multline}\label{eq:Mark3}
\GFkOD(z,z'')=-\dfrac{e^{ik(z''-z)}}{2ik}+\dfrac{e^{ik(z''-z)}}{4}\int_{-a}^{a} \Delta\hepsOD_k(z')dz'
\\
-\dfrac{ke^{ik(z''-z)}}{4}\sum_n \frac{\AnOD(k_s)\AnOD(k_s'')}{2(k-k_n)}\,,
\end{multline}
where $\AnOD$ is defined as the Fourier transform,
\be
\AnOD(k_s)=\int_{-a}^a e^{ik_sz'}\Delta\hepsOD_k(z')\EnOD(z')dz'\,,
 \label{Ani}
\ee
which in the case of a homogeneous slab treated here is calculated to be 
\be
\AnOD(k_s)=B_n\Delta\varepsilon\left[\dfrac{e^{i(k_s+\sqrt{\epsilon_s}k_n)z'}}{i(k_s+\sqrt{\epsilon_s}k_n)}+\dfrac{(-1)^ne^{i(k_s-\sqrt{\epsilon_s}k_n)z'}}{i(k_s-\sqrt{\epsilon_s}k_n)}\right]_{-a}^{a} 
 \label{Ani}
\ee
Interestingly in 1D we do not require the far field approximation to make the simplification of the Green's function required to bring the RSE Born approximation to the form of \Eq{eq:Mark3}. Hence in 
1D the RSE Born approximation is valid everywhere outside of the slab and not just in the far field. 

\section{Numerical Validation}\label{sec:application3D}

In this section we calculate the 1D GF outside of the homogeneous slab given by \Eq{unpslab} where $\epsilon_s=2.25$. We do this using the 
RSE Born approximation, analytically using 
boundary conditions, and also using the spectral GF for comparison. We find that the RSE Born approximation requires less basis states to reach 
a required accuracy than the spectral GF and unlike the 
spectral GF is convergent outside the system.

I calculate three types of GF in this section, an analytic GF, the GF of \Eq{ML4OD} and the GF of \Eq{eq:Mark3}. From these it is possible to 
use the formula derived in Ref.\cite{DoostPRA12, ArmitagePRA14}  for normal incident and
waveguide systems for the transmission $T(k,z')$,
\begin{equation}
T(k,z')=\left|2kG(z',-a;k)\right|^{2}\,. \label{Trans1}
\end{equation}

The analytic GF is found by solving Maxwell's wave equation in 1D with a source of plane waves while making use of Maxwell's boundary conditions.

The procedure used to select the basis of RSs for the RSE Born approximation calculation is analogous to that described in \Onlinecite{DoostPRA14} for the RSE perturbation theory. Namely, I 
choose the basis of RSs such that all RSs with  $|k_n|<k_{\rm max}(N)$ using a maximum wave vector $k_{\rm max}(N)$ chosen to select $N$ RSs.
\begin{figure}
\includegraphics*[width=\columnwidth]{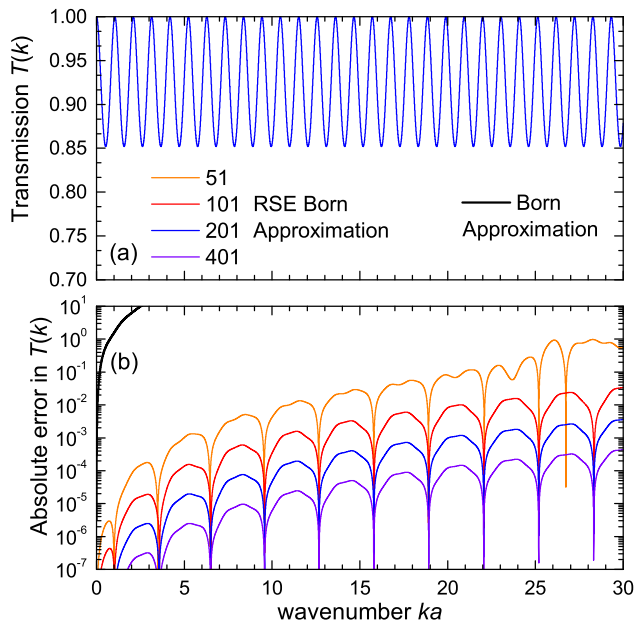}
\caption{(Color online)(a) Light transmission through the slab, \Eq{Trans1}. (b) Absolute error in the transmission calculated using the analytic form of $T(k,a)$  as comparison for the numerical values 
from the RSE Born approximation with $N=51, 101, 201, 401$ as labelled. For further comparison the standard Born approximation is also included.} \label{fig:Trans_vs_k_BA}
\end{figure}

From \Fig{fig:Trans_vs_k_BA} we can see that unlike the standard Born approximation the RSE Born approximation is valid over an arbitrarily wide range of $k$ depending only on the basis size $N$ used. 
Furthermore we see that as the basis size increases the RSE Born approximation converges to the exact solution. The absolute error in the RSE Born approximation is approximately reduced by an order 
of magnitude each time the basis size is doubled. Absolute errors of $10^{-7}-10^{-4}$ are seen in the $k$ range shown for basis size $N=401$.

\begin{figure}
\includegraphics*[width=\columnwidth]{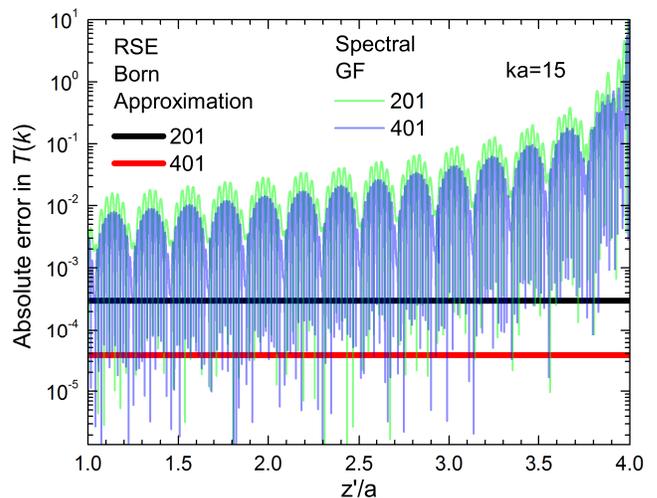}
\caption{Absolute error in transmission calculated using the analytic form of $T(k,z')$, \Eq{Trans1}, between $z=-a$ and $z=z'$ as comparison for the RSE Born approximation and the spectral GF at fixed $ka=15$. $N=201, 401$ as labelled.} \label{fig:Trans_vs_zp_BA_BRSE}
\end{figure}
From \Fig{fig:Trans_vs_zp_BA_BRSE} we can see that unlike the GF calculated with the spectral \Eq{ML4OD} the RSE Born approximation is stable over an arbitrarily wide range of $z'$, where $z'$ is the 
coordinate of the point of transmission to, depending only on the basis size $N$ used. The transmission calculated via the spectral GF is diverging with distance of the point of transmission from the slab, 
this suggests that outside the system the RSE spectral GF is not converging or is divergent. Furthermore we see that the RSE Born approximation requires fewer resonant states than the spectral GF method in 
order to produce a required absolute error, at all values of $z'$. Although these points were first noted by Ge {\it et al}.~\cite{Ge14} 
they were using the unstable normalisation leading to an incorrect GF
and so the results upon which they based their conclusions are not reliable.

\section{Summary}\label{sec:summary}
In this work we have seen the Born approximation mathematically rigorously extended to include terms which take into account the resonances of the scattering potential using the exact same method as 
\cite{Ge14} except with correctly normalised modes. Further I have made comparisons in 1D between scattering calculated with the spectral GF and the scattering calculated using the RSE Born approximation.
I have demonstrated that once the correct normalisation is used in the RSE Born approximation convergences towards the exact solution is obtained.
I have found that the RSE Born approximation for finding the full GF outside of the system is superior to the other spectral GF method considered in terms of convergence and accuracy when the correct 
normalisation of the RSs is used.

It is demonstrated in this paper that the 1D RSE Born approximation can be used in conjuncture with the RSE perturbation theory (to generate the normalised eigen-modes of planar systems with arbitrary dielectric profile) 
\cite{MuljarovEPL10,DoostPRA12, ArmitagePRA14, DoostARX15B} to offer a possible alternative to the scattering matrix method of Ref.\cite{CoIng}. In fact, given the superior efficiency of the RSE perturbation theory in comparison with FDTD and FEM
for weak perturbations demonstrated in Ref.\cite{DoostPRA14} it is likely that the RSE coupled with the RSE Born approximation will be an incredibly powerful scattering theory for weak scatterers.

I have now derived an analogous theory for general wave equations \cite{DoostARX15B}.

\appendix

\section{Derivation of alternative Green's function and completeness}
\label{App:L}

In order to simplify the RSE Born approximation and develop \Eq{normaliz} we require an appropriate spectral form of the GF which is different from the one already proven in the literature. To obtain this 
correct form I start with the GF valid inside the scatterer only, which I derived in Ref.\cite{DoostPRA13, DoostARX15B},
\be \GFk(\br,\br')=\sum_n \frac{\En(\br)\otimes\En(\br')}{2k_n(k-k_n)}\,.
\label{TUES_NIGHT1} \ee
Substituting \Eq{TUES_NIGHT1} in 
\be
- \nabla\times\nabla\times \GFk(\br,\br')+k^2\heps_k(\br)\GFk(\br,\br')=\hat{\mathbf{1}}\delta(\br-\br')\,,
 \label{GFequ0}
\ee
gives for $k\to \infty$,
\be \heps(\br)\sum_n\dfrac{(k+k_n)\En(\br)\otimes\En(\br')}{2k_n}=\hat{\mathbf{
1}}\delta(\br-\br')\,, \label{TUES_NIGHT2} \ee
since throughout the derivation in this appendix we are considering the limit where $k\to \infty$ at which $\heps_k(\br)=\heps(\br)$, i.e. the system is non-dispersive at high frequencies.

Convoluting \Eq{TUES_NIGHT2} with arbitrary finite functions and assuming the series are convergent we see that since $k\to \infty$ we have the sum 
rule,  
\be \sum_n \dfrac{\En(\br)\otimes\En(\br')}{2k_n}=0\,. \label{Sum-rule} \ee
Combining \Eq{TUES_NIGHT1} and \Eq{Sum-rule} yields
\be \GFk(\br,\br')=\sum_n \frac{\En(\br)\otimes\En(\br')}{2k(k-k_n)}\,.
\label{ML2C2} \ee
Combining \Eq{TUES_NIGHT2} and \Eq{Sum-rule} leads to the closure
relation
\be \dfrac{\heps(\br)}{2}\sum_n\En(\br)\otimes\En(\br')=\hat{\mathbf{
1}}\delta(\br-\br')\,, \label{Closure} \ee
which expresses the completeness of the RSs, so that any function can be written
as a superposition of RSs. If in the perturbed system some of the series are not convergent or are instead conditionally convergent then we will not arrive at the sum rule and completeness, 
in which case I expect that the RSE Born approximation will still give 
convergence to the exact solution but only if a valid spectral Green's function is used, such as \Eq{TUES_NIGHT1}. 

\section{RSE for dispersive systems}
\label{App:M}

Due to the problems with non-convergence of Schur factorisation for the generalised eigen-value problem of perturbing nano-spheres dispersively, the RSE in Ref.\cite{DoostARX15A} might tends to fail for non-symmetric perturbation when
more than typically $500$ basis states are used. This is an estimate based on the un-reported RSE failures for half and quarter sphere perturbations using the generalised eigen-value problem form of the RSE, tests which I 
carried out for Ref.\cite{DoostPRA14}. Therefore it is necessary to add linear
dispersion through a second stage perturbation, a perturbation to the possibly complex conductivity \cite{DoostARX15A}.

To make this perturbation consider the problem of a perturbation to the conductivity $-i{{\bf\Delta\hsigma}(\br)}/{k_{\mu}}$
\be
\label{me4D}
\nabla\times\nabla\times\bE_{\mu}(\br)=k_{\mu}^2\left[\heps_{k_{\mu}}(\br)-i\dfrac{{\bf\Delta\hsigma}(\br)}{k_{\mu}}\right]\bE_{\mu}(\br)\,.
\ee
$\heps_{k_{\mu}}$ could have in principle any dispersion for which the eigen-modes can be normalised, and which becomes non-dispersive in the limit of high frequency in order to make the sum rule for the GF. 

In this Appendices Greek index letters denote perturbed modes and British (English) lower case index letters denote unperturbed modes.

Since
\be 
\bE_{\mu}(\br)=ik_{\mu}\int\GFkm(\br,\br'){{\bf\Delta\hsigma}(\br')}\bE_{\mu}(\br') d\br'\,,
\ee
where $\GFkm(\br,\br')$ is given by \Eq{ML2C2} and by \Eq{Closure}
\be
\label{latenight}
\bE_{\mu}(\br)=\sum_n b_{n\mu}\bE_{n}(\br)\,,
\ee
where in \Eq{latenight} $\bE_{n}$ and $k_n$ correspond to the unperturbed modes of \Eq{me3D}, then following the derivation method of Ref.\cite{NMuljarovEPL101,NMuljarovEPL102}
\be
2b_{n\mu}k_\mu=\sum_a\left(i{\bf S}_{na}+2\delta_{na}k_a\right)b_{a\mu}\,,
\ee
where
\be
{\bf S}_{na}=\int\bE_{n}(\br)\cdot {\bf\Delta\hsigma}(\br)\bE_{a}(\br)d\br\,,
\ee
which can be solved for the eigen-modes $\bE_{\mu}$ and eigen-values $k_{\mu}$ of the perturbed problem.

For examples of fitting dispersion linear in wavelength to the dispersion of real materials for the purposes of RSE perturbation theory please see my Ref.\cite{DoostARX15A} where a linear dispersive RSE is presented in terms of a 
generalised eigen-value problem.

The perturbation to $\heps_{k_{\mu}}$ can also be non-dispersive without resorting to generalised eigen-value problems, as treated in my Ref.\cite{DoostPRA14}. 
To elaborate further on this point consider the problem of the non-dispersive perturbation $\Delta\heps(\br)$
\be
\label{me5D}
\nabla\times\nabla\times\bE_{\mu}(\br)=k_{\mu}^2\left[\heps_{k_{\mu}}(\br)+\Delta\heps(\br)\right]\bE_{\mu}(\br)\,.
\ee
Again $\heps_{k_{\mu}}$ is dispersive as in \Eq{me4D}. \Eq{me5D} is solved by \cite{DoostPRA14}
\be
\label{me6D}
\sum_a\left(\dfrac{\delta_{na}}{k_a}+\dfrac{{\bf V}_{na}}{2\sqrt{k_n}\sqrt{k_a}}\right)c_{a\mu}=\dfrac{1}{k_\mu}c_{n\mu}\,,
\ee
where
\be
{\bf V}_{na}=\int\bE_{n}(\br)\cdot {\bf\Delta\heps}(\br)\bE_{a}(\br)d\br
\ee
and $b_{n\mu}\sqrt{k_n}=c_{n\mu}$. Please note that it is very important to be consistent with the signs of $\sqrt{k_b}$ in the matrix elements of \Eq{me6D}.

Using the linear eigen-value approach outlined here it might be possible to treat an unperturbed Drude-Lorentz gold sphere with a non-dispersive 
shell, and perturb away the non-dispersive shell leaving in its place biological particles to be 
sensed as a perturbation. All perturbations must be within the boundaries of the unperturbed system due to convergence of the GF, see 
\Fig{fig:Trans_vs_zp_BA_BRSE}.

For a discussion of the eigen-functions of Maxwell's equations in spherical coordinates please see \cite{Coilin}.

\section{Kristensen normalisation}
\label{App:SF}

In order for the normalisation of Kristensen {\it et al} to be correct it must be consistent with my \Eq{normaliz}, specifically the surface term $S^{RSE}$ in \Eq{normaliz} must be mathematically equivalent with Kristensen's 
surface term $S^{KS}$. 
Hence it should be that $S^{RSE}=S^{KS}$, where from \cite{MuljarovEPL10} we have,
\be
\label{GB}
S^{RSE}=+\frac{1}{2k^2_n}\oint_{S_R}\!\!\!\! dS \left[\En\!\cdot\!\frac{\partial}{\partial r}r\frac{\partial\En}{\partial r}-r\!\left(\frac{\partial \En}{\partial r}\right)^{\!2}\! \right]\,,
\ee
and from \Eq{norm2} we have
\be
\label{MIG17}
S^{KS}=\lim_{{V}\to\infty}\frac{i}{2k_n} \oint_{S_{V}} \En^2 (\br) dS\,,
\ee
However considering the RSE Born approximation in 3D, specifically to ensure outgoing BCs \Eq{eq:Mark3b} in particular driven by (convoluted with) a current 
vanishing proportional to $(k-k_n)$ as $k\rightarrow k_n$ 
so $\mathbf{E(\br)}\rightarrow\En(\br)$, we know that 
\be
\label{GBA}
\lim_{r\rightarrow\infty}\En(\br)={\bf f}(\theta,\phi)\dfrac{e^{ik_nr}}{r}\,,
\ee
where $\br=(r,\theta,\phi)$ in spherical polar coordinates and ${\bf f}(\theta,\phi)\propto\An(\hat{\br}k_n)$, 
then substituting \Eq{GBA} into \Eq{GB} and \Eq{MIG17} and equating $S^{RSE}$ and $S^{KS}$ 
gives,
\be
\label{GBA2}
\lim_{r\rightarrow\infty}S^{RSE}=\lim_{r\rightarrow\infty}S^{KS}=\dfrac{ie^{2ik_nr}}{2k_nr^2}\oint_{S_{V}} {\bf f}^2(\theta,\phi) dS\,,
\ee
a logically valid statement, therefore, $S^{RSE}= S^{KS}$ when $r\rightarrow\infty$ and so the normalisation of Kristensen {\it et al} 
is not wrong as it is stated. In the RSE perturbation theory letting  $r\rightarrow\infty$ in the normalisation 
of perturbed modes introduces huge errors because of the blow up of the RS mode fields far from the system causing blow up of error. 
By \Eq{GBA2}, as $r$ grows one is essentially subtracting an exponentially growing surface term from an exponentially growing
volume term to get the constant $1+\delta_{k_n,0}$ for normalisation, this leads to large numerical errors for low-Q (leaky) modes \cite{KristensenOL12}.
An inherent source of instability is remaining dependence of ${\bf f}(\theta,\phi)$ on $r$ due to the use of finite $r$.

In 1D Kristensen {\it et al} normalisation is actually correct for any finite $r$ that includes the system inhomogeneity \cite{MuljarovEPL10}.

The remaining problems with the Kristensen {\it et al} normalisation is that it is missing $k=0$ modes, therefore it is incomplete, 
and hence incorrect. Also it does not have the conditions on $\heps_k$ which should be the same as
for my normalisation. That the RSs can be written in the form of \Eq{GBA} aids the solution of the inverse scattering problem \cite{DoostARX15B}.

As an aside, since
\be
\label{GBA3}
{\bf f}(\theta,\phi)=\dfrac{k^2_n}{4\pi}\An(\hat{\br}k_n)\,,
\ee
it is theoretically possible to make calculations of the potential from the emission (decay) via
fast inverse Fourier transform methods upon the set of $\An(\hat{\br}k_n)$ especially if the 
potentials of interest are rotating about a fixed axis so we know their orientation to some
extent such as occurs for decaying magnetic nuclei as part of a non-magnetic crystalline
compound placed inside a NMR (nuclear-magnetic-resonance) machine. Because $k_n$ are
discrete values these inverse Fourier methods might have to be used self-consistently in conjuncture
with the RSE perturbation theory and the values of $k_n$. This is a highly speculative aside and
might be a possible topic for future research.

As another aside, close to a sharp resonance $k\approx k_n$ scattering from the potential is
dominated by a single resonance and so the scattered $\mathbf{E}$-field $\mathbf{E^{scattered}}$ is approximately
\be
\label{GBA4}
\lim_{r\rightarrow\infty}\mathbf{E^{scattered}}(\br, k)\approx C\An(\hat{\br}k)\dfrac{e^{ikr}}{r}\,,
\ee
again $\An(\hat{\br}k)$ can be partially inverse Fourier transformed with respect to angle to find
information about the internal structure of the potential. C is some constant.
This argument assumed elastic scattering, which is a valid assumption even for such things
as neutron-nucleus scattering provided that the neutron energies are high enough, at low
energies inelastic scattering causes deviations from the RSE Born approximation model,
these deviations in scattering caused by inelasticity give the fission or absorption cross-sections. An 
RSE for Schr\"odinger's equation is given in \cite{DoostARX15B}.

For 2D systems, by following similar arguments as here we arrive at (in the notation of \cite{DoostARX15B})
\be
\label{GBA5}
\lim_{\rho\rightarrow\infty}\En(\brd)=-\omega^2_n\dfrac{Q}{\sqrt{\rho\pi}}\An(\hat{\brd}k_n)e^{ik_n\rho}\,,
\ee
and the arguments with regards to normalisation in this Appendix C still hold except that now

\bea &&\!\!\!\!\!\!\!1+\delta_{k_n,0}=\int_A\En\cdot\dfrac{\partial(\omega^2\heps_\omega)}{\partial(\omega^2)}\bigg|_{\omega=\omega_n}\En d{\brd} \label{normaliz222} \\
 &&\!\!\!\!\!\!\! +\frac{R}{2k_n^2}\int_0^{2\pi}d\theta \left[\En\cdot\frac{\partial\En}{\partial\rho}+\rho \En\cdot\frac{\partial^2\En}{\partial \rho^2}-\rho\left(\frac{\partial\En}{\partial\rho}\right)^2\right]_{\rho=R}\,.\nonumber \eea

\end{document}